\documentclass[twocolumn,english,aps,prl,superscriptaddress,showpacs]{revtex4-1}
\usepackage[latin9]{inputenc}
\usepackage{amsmath}
\usepackage{amssymb}
\usepackage{graphicx}

\usepackage{babel}

\begin{document}

\title{Universal Bose Gases Near Resonance:
A Rigorous Solution}

\author{Shao-Jian Jiang}
\affiliation{Beijing National Laboratory for Condensed Matter Physics,
	Institute of Physics, Chinese Academy of Sciences, Beijing 100190,
	China}
\affiliation{Department of Physics and Astronomy, University of British
	Columbia, Vancouver V6T 1Z1, Canada}
\author{Wu-Ming Liu}
\affiliation{Beijing National Laboratory for Condensed Matter Physics,
	Institute of Physics, Chinese Academy of Sciences, Beijing 100190,
	China}
\author{Gordon W. Semenoff}
\affiliation{Department of Physics and Astronomy, University of British
	Columbia, Vancouver V6T 1Z1, Canada}
\author{Fei Zhou}
\affiliation{Department of Physics and Astronomy, University of British
	Columbia, Vancouver V6T 1Z1, Canada}
\affiliation{ICQS, Institute of Physics, Chinese Academy of Sciences,
	Beijing 100190, China}

\date{July 4, 2013}
\begin{abstract}
We obtain a rigorous solution of universal Bose gases near resonance and offer an answer to one of the long-standing challenges of quantum gases at large scattering lengths, where the standard dilute theory breaks down. 
The solution was obtained by using an $\epsilon$ expansion near four spatial
dimension. In dimension $ d=4-\epsilon$, the
chemical potential of Bose gases near resonance is shown to approach the
universal value $\epsilon^{\frac{2}{4-\epsilon}} \epsilon_F \sqrt{\frac{2
		}{3}} (1+0.474 \epsilon - i 1.217
\epsilon + \cdots )$, where $\epsilon_F$ is the Fermi energy defined for
a Fermi gas of density $n$, and the condensation fraction is equal to  $\frac{2}{3}(1 + 0.0877
\epsilon + \cdots)$. We also discuss the implications on ultra-cold gases in physical dimensions.
\end{abstract}
\maketitle
In recent applications of Feshbach resonances, a few cold-atom labs have explored quantum gases of scattering atoms at 
large positive scattering lengths~\cite{Papp08,Pollack09,Jo09,Navon11,Wild12,Ha13,Fletcher13,Inouye98}, 
a subject that is beyond the known dilute gas
theories for weakly scattering atoms and hence extremely poorly understood.
This limit of large scattering lengths with almost no formation of dimers has been called the {\it upper branch} Feshbach resonance, 
as opposed to the {\it lower branch} that is best known for its intimate
connection to the BCS-BEC crossover physics~\cite{Ketterle07,Chin10,Leggett80,Eagles69}. The upper branch physics is an excellent example of the unique complexities of resonant cold gases.
It puts one of the challenges in quantum many-body 
physics, that is quantum gases at large scattering lengths, under the spotlight. 
Although the physics of lower branch unitary Fermi gases is described by the BCS pairing wave functions and
hence has been quite well understood after years of intensive research, the situation for the upper branch resonant gases has been much less encouraging.
For Bose gases, the dilute gas theory was developed more than half a century ago~\cite{Bogoliubov47,Lee57,Beliaev58,Wu59,Sawada59,Pines59}. 
The latest attempt to include higher order corrections in the dilute limit was made 
a while ago to address the effect of Efimov trimers~\cite{Braaten02,Efimov70}.
However, the existing dilute gas theories are obviously not applicable
in the limit of large scattering lengths.

In this Letter, we take a new approach instead of launching another frontal attack on this puzzling limit of large scattering lengths.
It is based on the {\it $\epsilon$} expansion 
near four spatial dimension (4D) and it provides a rigorous solution to Bose gases near resonance.
A few years ago, Yang had discussed possible extensions of pseudo-potential in
dimensions higher than three and illustrated that the pseudo-potential
further depends on the details of interactions at short distances and
becomes ill-defined in high dimensions~\cite{Yang}.
The peculiar feature of resonance scattering in 4D on the other hand was first noticed by Nussinov and Nussinov who found that the wave function of  
two scattering atoms is concentrated at short distance~\cite{Nussinov}. 
Later, Nishida and Son had constructed a successful renormalization scheme to evaluate the effective
potential or the energy density for the paired fermions in $4-\epsilon$ dimensions~\cite{Nishida06} (see other schemes in
Ref.~\cite{Nikolic07,Veillette07}).

The main motivations of our studies here are at least two-fold. 
Since Feshbach resonances were applied to study many-body physics in laboratories, there have
been a few attempts of developing non-perturbative approaches to near-resonance physics.
One of the exciting directions is to utilize the ultra-violet properties of the momentum distribution function
to establish an exact relation for the energy~\cite{Tan08,Tan08b,Punk07}. 
However, unlike in Fermi gases where a universal contact parameter can be introduced~\cite{Tan08},
for Bose gases additional non-universal regularization had to be carried out~\cite{Werner10,Braaten11}; its implications need to be further examined. 
The other direction that has been quite intensively pursued for bosons is to directly
evaluate the effective potential for the condensed atomic field~\cite{Borzov12,Zhou13,Mashayekhi13}. 
This approach is equivalent to applying scale dependent interaction constants in energy calculations with an emphasis 
on the infrared physics.
It takes into account the varying magnitudes and signs of the
running coupling constants over low energy scales.
Although the self-consistent framework in these attempts is exact, 
in the absence of a controllable expansion parameter
the self-consistency was practically implemented
via including correlations in up to three-body channels.
Quantitatively, these theories are approximate and {\it a posteriori}.
They should be tested in either experiments or more sophisticated Monte Carlo simulations both of which are in infancy as far as the 
upper branch resonant Bose gases are concerned~\cite{Giorgini99}.

Given the current status of theories and experiments, 
a rigorous solution, even though in higher spatial dimensions, can provide
enormous insight and even constraints on correct theories of quantum gases
in physical dimensions. It can serve as an important benchmark for future theoretical attempts to understand resonant gases of scattering atoms~\cite{Qi08}.
Furthermore, there has been evidence that 3D Bose gases are not universally characterized by two-body scattering parameters because of the ultraviolet physics related to Efimov
states~\cite{Efimov70,Braaten02,Werner10,Borzov12}. One might ask whether there exist universal Bose gases in other spatial dimensions  or other universalities of Bose gases.
The $\epsilon$ expansion near 4D in this paper provides a definite answer to this question of universality.

Recall that in 3D Bose gases, the Lee-Huang-Yang (LHY) correction is purely a collective
effect~\cite{Lee57} and gets contributions from all N-body effects with
$N=3,4,5...$; this is one of the main reasons why higher order effects are very difficult to thoroughly examine.
$\epsilon$ expansion provides an effective way to systematically study N-body contributions near resonance. This can be understood by considering 
the Born-Oppenheimer potential of two non-interacting heavy bosons
resonantly scattered by a light one~\cite{MacNeill11,Zinner13}. 
Near 4D, the ground state energy of the three bosons
with two heavy ones fixed at distance $|{\bf R}|$ apart can be easily estimated. One can then show that 
the Born-Oppenheimer potential between two heavy bosons mediated by a light
one scales as $\epsilon |{\bf R}|^{-2}$ and is suppressed by an extra
$\epsilon$ factor in $d= 4 - \epsilon$. For the quantum gas under consideration, this implies the contribution from
$N$-body forces with $N > 2$ should be systematically expandable in terms of {\it $\epsilon$}. This insight is particularly useful for our analysis.

We shall apply the $\epsilon$ expansion near 4D to the
upper branch bosons. We will implement it with two important new elements. First,
since we are dealing with an upper branch, 
in principle the energy density 
has an imaginary part indicating a coupling to the lower
branch. This shows up as a higher order effect in the dilute gas theories
while, in the $\epsilon$ expansion, it
appears as a leading order correction to the energy density near resonance
and it therefore must be included in our discussion. Second, the non-interacting Bose
gases are infinitely compressible and
therefore even in the dilute limit the energy density as a function of
scattering length contains terms with fractional powers of the scattering
length. This issue can be effectively dealt with by further combining
the method of $\epsilon$ expansion with self-consistent equations.

A condensate with a contact interaction can be described by
\begin{eqnarray}
&&H - \mu \sum_{\bf k} b_{\bf k}^\dagger b_{\bf  k}  \nonumber \\
&=&\sum_{\bf k} (\epsilon_{\bf k} -\mu) b_{\bf k}^\dagger b_{\bf  k}
+ 2 U_0 n_0 \sum_{\bf k} b^\dagger_{\bf  k} b_{\bf  k}
+\frac{1}{2} U_0 n_0\sum_{\bf k}
b^\dagger_{\bf  k}
b^\dagger_{-\bf  k}
\nonumber \\
&+&
\frac{1}{2}U_0 n_0 \sum_{\bf k} b_{\bf  k}b_{-\bf  k}
+\frac{U_0}{\sqrt{\Omega}}\sqrt{n_0}
\sum_{{\bf k'},{\bf q}} b^\dagger_{\bf q} b_{\bf  k'+\frac{\bf q}{2}}
b_{-\bf  k'+\frac{\bf q}{2}}+h.c.
\nonumber \\
&+&\frac{U_0}{2\Omega} \sum_{{\bf k}, {\bf k'},{\bf q}} b^\dagger_{\bf k+\frac{\bf q}{2}} b^\dagger_{-\bf  k+\frac{\bf q}{2}} b_{\bf  k'+\frac{\bf q}{2}}
b_{-\bf  k'+\frac{\bf q}{2}}+h.c.
\end{eqnarray}
where $\epsilon_{\bf k}\!=\!\hbar^2 {\bf k}^2/(2m)$; $\hbar$ is the reduced
Plank constant and $m$ is the mass of a single atom. We will set
$\hbar$ and $m$ to be unity from here on.
The sum is over non-zero momentum states.
$U_0$ is the strength of the contact interaction related to the
renormalized 2-body coupling constant $g_2$ via $U_0^{-1}\!=\!g_2^{-1}\! -\!\Omega^{-1}\! \sum_{\bf k}\! (2\epsilon_{\bf
	k})^{-1}$, $\Omega$ is the volume, and $g_2$ is determined by the size
of the 2-body bound state $\lambda_B$ 
\begin{equation}
	g_2 = \frac{ -(4 \pi) ^{2-\epsilon/2}}{ \Gamma(\frac{\epsilon}{2}-1
		)} \lambda_B^{2-\epsilon}, \Gamma \left( \frac{\epsilon}{2}
		- 1\right) \xrightarrow[]{\epsilon \rightarrow 0} - \frac{2}{\epsilon}
\end{equation}
in $4-\epsilon$ dimensions, where $\Gamma$ is the gamma function. 
$n_0$ is the number density of condensed atoms and $\mu$ is the
chemical potential of non-condensed particles, both of which are functions
of $\lambda_B$ and $\epsilon$ and are to be determined self-consistently.

The energy density for a fixed $n_0$ and $\mu$ can be obtained as $E(n_0,\mu)$;
then the following set of self-consistent equations can be applied to
study the chemical potential for a gas with total number density $n$,
\begin{eqnarray}
\mu_c (n_0, \mu) &=&\frac{\partial{E(n_0,\mu)}}{\partial n_0},
n=n_0 -\frac{\partial{E(n_0,\mu)}}{\partial \mu}, \nonumber \\
\mu &=& \mu_c(n_0, \mu),
\label{SCD}
\end{eqnarray}
where $\mu_c$ is the chemical potential for the condensed atoms.
In equilibrium, $\mu_c$ has to be equal to $\mu$, the chemical potential of non-condensed atoms as indicated in Eq.~\eqref{SCD}.
Calculations of $E(n_0,\mu)$ are carried out diagrammatically using the
standard effective field theory method~\cite{Beliaev58,Coleman73}.
This quantity in 2D and 3D was studied in Ref.~\cite{Borzov12,Mashayekhi13}.
The general structure of $E(n_0,\mu)$ is given below. Its $\epsilon$ dependence is shown explicitly.
\begin{eqnarray}
E(n_0, \mu) \!
= \!
 \frac{g_2 n_0^2}{2}\! \sum_{N \ge 2} (2 g_2 n_0 \lambda_B^2)^{N-2} 
A^{(N)}(k_\mu \lambda_B,\epsilon),
\label{ED}
\end{eqnarray}
where $A^{(N)}(k_\mu\lambda_{B},\epsilon)$ represent the contributions from
the renormalized $N$-body forces, and $k_{\mu} = \sqrt{2  \mu}$.

From the point of view of running coupling constants~\cite{Zhou13},
the healing length $\xi=1/k_\mu$ is a crucial length scale which separates the short distance few-body physics 
controlled by the renormalization flow of coupling constants from the long wavelength hydrodynamic regime of cold gases where collective effects dominate.
At the healing length, the usual renormalization flow generated under scale
transformation is subject to a boundary condition due to a thermodynamic constraint. Alternatively, one states that the chemical potential is dictated by 
the running coupling constants at the scale of healing length, which leads to a self-consistent equation.
This is also fully reflected in Eq.~\eqref{SCD} and Eq.~\eqref{ED}, where the running coupling constants $A^{(N)}(k_\mu\lambda_{B},\epsilon)$, 
$N=2,3...$ defined at a pre-assumed healing length $\xi= 1/k_\mu$ are
further applied to evaluate the chemical potential.

\begin{figure}[t]
	\includegraphics[width=2.5in]{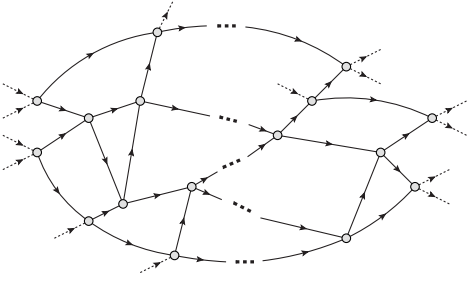}
	\caption{An $N$-body $L$-loop diagram with $N$ incoming and outgoing condensed lines
		(dashed); $L$ is the number of loops formed by propagators (solid lines). Each diagram has $N\!+\!(\!L\!-\!1\!)$ vertices. Near resonance, its contribution to the
		chemical potential is proportional to
		$\epsilon^{\frac{2}{4-\epsilon}}\epsilon^L$ independent of $N$.}
	\label{N_L}
\end{figure}

We have carried out a thorough study on these renormalized forces and shall report our results here.
Detailed derivations will be published in a follow-up technical article.
The energy density (in unit of the Hartree-Fock energy $ g_2 n_0^2/2
\approx 4
\pi^2\epsilon  \lambda_B^{2-\epsilon} n_0^2$) turns out to be a function of two dimensionless parameters,
$\epsilon$ and $n_0 \lambda_{B}^{4-\epsilon}$.
The contribution to $A^{(N)}$ is further specified by coefficients
$a^{(N)}_{L}$ , $b^{(N)}_{L}$ and $c_L^{(N)}$ (see
Ref.~\cite{dilute-limit} and below) with $L$ standing for the number of loops in the diagrams involved as illustrated in Fig.~\ref{N_L}.
The asymptotic behaviors of $A^{(N)}(k_\mu\lambda_{B},\epsilon)$ when
$\epsilon$ becomes zero very much depend on the self-consistent parameter, $k_\mu\lambda_B$, which we are
now turning to. 
In 3D, the dilute limit is defined as the limit where the number of atoms
within the volume defined by the size of dimers, $\lambda_{B}$
(or scattering length $a=\lambda_{B}$), is much less than one i.e.
$n\lambda_B^3 \ll 1$. 
Therefore, we define the gas parameter in $4-\epsilon$ dimension simply to be
$n\lambda_{B}^{4-\epsilon}$ and the dilute limit is
$n\lambda_{B}^{4-\epsilon} \ll 1$.
We shall discuss this limit first.
There, the dominating contribution to the energy density in this limit is the
Hartree-Fock energy  
$g_2 n_0^2/2$ and the corresponding chemical potential is $g_2 n_0$. The
self-energy in this limit is $\Sigma = 2 g_2 n_0 = 2 \mu$, and the relevant
momentum scale is $k_{\mu} = \sqrt{2 \mu}$.
The leading correction is purely from irreducible 2-body contributions
which are of a form  
\begin{eqnarray}
A^{(2)}
= 1+ (k_{\mu} \lambda_B)^{(2-\epsilon)}+...
 \label{dilute_2N}
\end{eqnarray}
In the dilute limit, the second term in Eq.~\eqref{dilute_2N}  scales as
$(\epsilon \lambda_B^{4-\epsilon} n_0)^{1-\epsilon/2}$ and yields the most
dominating correction to the Hartree-Fock energy (first term), which
is an analogue of LHY effect in 3D; the other terms that are not shown
explicitly in Eq.~\eqref{dilute_2N} are further
suppressed by higher powers of $ (\epsilon 
\lambda_B^{4-\epsilon} n_0 )^{1-\epsilon/2}$.
The next order correction contains an additional power of $\epsilon$ and
has both real and imaginary parts.  The real
part  is from the leading order 
$N=4,6,...$ terms, and imaginary part from the leading order terms with
$N=3$~\cite{dilute-limit}.

One can compute the energy density and then solve Eq.~\eqref{SCD} for the chemical potential  perturbatively in the low density limit. The result is
\begin{eqnarray}
Im \mu\! &=&\!- \frac{ -(4 \pi) ^{2-\frac{\epsilon}{2}}}{
		\Gamma(\frac{\epsilon}{2}-1 )} \lambda_B^{2-\epsilon} n_0
 \! \left (\!\frac{ -2(4 \pi) ^{2-\frac{\epsilon}{2}}}{
				\Gamma(\frac{\epsilon}{2}-1 )}
			\lambda_B^{4-\epsilon}  n_0 \!
		\right)\! \epsilon \frac{3\pi}{2}\! +\! ... \nonumber \\
	Re \mu \!& =&\! \frac{ -(4 \pi) ^{2- \frac{\epsilon}{2}}}{
		\Gamma(\frac{\epsilon}{2}-1 )} \lambda_B^{2-\epsilon} n_0
	\!
	\left  \{\! 1\! +\!\left (\!\frac{ -2(4 \pi) ^{2-\frac{\epsilon}{2}}}{
				\Gamma(\frac{\epsilon}{2}-1 )}
			\lambda_B^{4-\epsilon}  n_0\!
		\right)^{1-\frac{\epsilon}{2}} \right .\nonumber \\
	       &&\times \left.	\left(2+ \epsilon \left( \frac{1}{2}
				\ln 4 - \frac{5}{4}\right)\right)+...\right
	\}. 
\label{chem}
\end{eqnarray}
The dimensionless parameter $\eta\!=\!2 g_2 n_0 \lambda_B^2\!\approx\!16 \pi^2 \epsilon  \lambda_B^{4-\epsilon} n_0$ appears naturally in our result because 
it defines the ratio between the Hartree-Fock chemical potential $g_2 n_0
\sim \epsilon \lambda_{B}^{2-\epsilon} n_0$ and the molecular binding
energy $1/\lambda_B^2$, which is a measure of the effective interaction
strength. When extrapolated to the limit $\epsilon=1$, the leading correction scales as $\sqrt{n\lambda_B^3}$ resembling the
LHY result in 3D.

Now we turn to the most interesting limit
where $\eta$ is of order of unity or even larger.
When $k_{\mu} \lambda_B \gg 1$, one can easily show that
$A^{(N)}$ is still an analytical function of $\epsilon$ and contains no
singular terms (See Fig.~\ref{N_L}) . For instance
for $N>2$,
\begin{eqnarray}
	A^{( \! N \! )}(k_\mu\lambda_{B}\! \rightarrow \! \infty,
	\! \epsilon)\!= \! \sum_{L=1}^{\infty} \! b^{( \! N \! )}_L \!
	\epsilon^{L} \! (k_{\mu} \lambda_B\!)^{-4N+6 + \epsilon (\!N-1\!) }. &&
\end{eqnarray}
Since $L=2,3,4...$-loop diagrams contain higher powers of
$\epsilon$ and become negligible when approaching 4D,
the dominating contributions are simply $L=1$-loop, $N=3,4,5...$-body diagrams
that contain both real and imaginary parts; the imaginary parts
represent the  N-body recombination processes.
This aspect is unique near 4 spatial dimension and provides a
systematic way to sum up contributions even though the quantum gas is near
resonance or $n\lambda_{B}^{4-\epsilon} \epsilon \gg 1$. 
In the linear order of $\epsilon$, the self-consistent equations in Eqs.~(\ref{SCD}) and (\ref{ED}) can be cast in a simple form 
\cite{Energy_at_Res},
\begin{eqnarray}
	\frac{n}{n_0}\!&=&\! 1\!+\! Re\! \left[\!\frac{\eta (-i
				Z)^{-\epsilon}}{2(1\!-\!(-iZ)^{2-\epsilon})^2}
			\frac{2\!-\!\epsilon}{2}\! +\! \epsilon d\!
		\left( \frac{Z^2}{\sqrt{\eta}},
			\frac{1}{\sqrt{\eta}}\right)\right]+... \nonumber \\
	Z^2\!& =&\! \frac{\eta}{1-(-i Z)^{2-\epsilon}} +  \epsilon
	\eta^{\frac{2}{4-\epsilon}}
		f\left(\frac{Z^2}{\sqrt{\eta}},\frac{1}{\sqrt{\eta}}\right)
		+...
\label{self-cons}
\end{eqnarray}
where $Z=k_\mu\lambda_{B}$. $d(x,y)$ and $f(x,y)$ are two dimensionless
functions defined
as,
\begin{widetext}
\begin{eqnarray}
	d(x,y) &\equiv& \frac{8 \pi^2}{4 (x+y)^2} \left\{ i \int \frac{d \nu}{2\pi} \frac{d^4 q}{(2 \pi)^4}
 \frac{ h_+(x,y)
		h_-(x,y)[ \frac{4}{x+y} +  h_+(x,y) + h_-(x,y) + 2
		l_+^2(x,y) h_+(x,y) + 2 l_-^2(x,y)
		h_-(x,y)]}{1- \frac{1}{4(x+y)^2}  h_+(x,y)
			h_-(x,y)} \right. \nonumber \\
		&&\left. - \frac{4}{x+y} \int \frac{d^4 q}{(2 \pi)^4} \frac{1}{x -q^2+ i
			\delta} - \int \frac{d^4 q}{(2 \pi)^4} \frac{2}{(x -q^2 + i
			\delta)^2} \right\},\nonumber
	\label{dfunc}
\end{eqnarray}
	\begin{eqnarray}
	f(x,y) & \equiv & 
        \frac{8\pi^2}{(x+y)^2}\left\{ i \int \frac{d \nu}{2\pi}
		\frac{d^4
		q}{(2\pi)^4} \frac{h_+(x,y)h_-(x,y)[1+
		\frac{1}{2}l_+(x,y)h_+(x,y)+\frac{1}{2}l_-(x,y) h_-(x,y)]}{1-
		\frac{1}{4(x+y)^2}h_+(x,y)h_-(x,y)}
	 - \int \frac{d^4
		q}{(2\pi)^4} \frac{1}{x - q^2 +i
		\delta}\right\}, \nonumber
\label{ffunc}
\end{eqnarray}
\end{widetext}
where $l(\nu,q;x,y)\! =\! \left(\!\nu\! -\! \frac{q^2}{4} 
	\!+\!x\! +\!y\! +\! i \delta \! \right)^{-1}$, $h(\nu,q;x,y)\! =\!
\left(\!\nu\! -\! \frac{q^2}{2}\! +\! \frac{x}{2}\! -\! l(\nu,q;x,y)\! +\! i
	\delta\!\right)^{-1}$, $h_{\pm} (x,y) = h(\pm \nu, q;x,y)$, and
$l_{\pm} (x,y) = l(\pm \nu, q;x,y)$ .

It is important to note that the right hand side of Eq.~\eqref{self-cons} 
is a function of the self-consistent variable $Z$ and two dimensionless parameters: $\eta$ and $\epsilon$.
When $\eta$ is small, the solution reproduces the dilute limit result. 
Near resonance when $\eta
\rightarrow \infty$, Eq.~\eqref{self-cons} yields a solution that is  universal, independent of ultraviolet physics, 
\begin{eqnarray}
	\mu &=&\epsilon^{\frac{2}{4-\epsilon}} \epsilon_F \sqrt{\frac{2
		}{3}} (1+0.474 \epsilon - i 1.217
\epsilon + \cdots ), \nonumber \\
n_0 &=&  \frac{2}{3}n(1 + 0.0877 \epsilon + \cdots).
\label{universal}
\end{eqnarray}
Here one can see that indeed $k_{\mu} \lambda_B \sim \eta^{1/4} \gg1$.
The leading terms in Eq.~\eqref{universal} are fully dictated by the renormalized
two-body interactions.
N-body interactions with $N > 2$ only contribute to the
corrections proportional to $\epsilon$.
Eq.~\eqref{universal} also indicates that the chemical potential of a unitary Bose gas is
proportional to $\epsilon^{1/2}$ near 4D and
its life time which is inversely proportional to the imaginary part of the chemical potential scales as $\epsilon^{-3/2}$. 
Although $\mu$ vanishes as $\epsilon$ goes to zero, it scales as
$\epsilon^{1/2}$ instead of  $\epsilon$ as in the dilute limit, indicating a
strongly interacting regime. The corresponding condensation
fraction near 4D appears to approach the value of $2/3$.
By contrast, in 3D, the chemical potential further depends on a
non-universal three-body ultraviolet momentum scale~\cite{Braaten02,Borzov12}.

Very recently, a few theoretical attempts have been made to understand upper branch Bose gases in 2D and 3D via applying a 
single-parameter scaling approach to the running coupling constants~\cite{Borzov12,Zhou13,Mashayekhi13}. The main
intention there was to provide a simple theoretical framework on upper branch Bose gases, analogous to the BCS-BEC crossover theory of unitary Fermi gases. 
It was illustrated that the chemical potential reaches a maximum at a critical scattering length or density and Bose gases are nearly fermionized
before an onset of many-body instability sets in and the compressibility becomes negative~\cite{Pekker11,Pilati05}.
The predicted correlation between the instability and occurrence of fermionization near the maximum still needs vindication in experiments.
In 4D, the Bose gases are more stable and even very close to the resonance
the life time (scales as $\epsilon^{-3/2}$) 
is much longer than the many-body time scale defined by the chemical
potential (scales as $\epsilon^{-1/2}$).
The main reason for this difference between 4D and 3D or 2D is that three-, four-body processes etc. become strongly suppressed as $\epsilon$
approaches zero. Consequently, the mean-field shift of the dimer binding energy
which results in instabilities at finite scattering lengths in 2D and 3D~\cite{Zhou13}, is expected to be vanishingly small 
near 4D.

Despite of this difference in the life time, Eq.~\eqref{universal} still offers unique and valuable implications about Bose gases in 
physical dimensions.   
For instance when extrapolated to the limit of $\epsilon=1$ or 3D,
Eq.~\eqref{universal} does imply that the chemical potential is of order of the Fermi energy $\epsilon_F$ and
so Bose gases are nearly fermionized. This is in agreement with the previous numerical evidence in 3D~\cite{Song09,Cowell02,Diederix11} as well as 
the lower bound of chemical potentials measured in experiments~\cite{Navon11}.
Furthermore, the extrapolation also indicates that in 3D
the quantum depletion fraction or the fraction of non-condensed atoms is $0.275$,
surprisingly close to the value of $0.27$ obtained in Ref.~\cite{Borzov12}.
More importantly, it is mainly from the two-body channel ($0.333$) while the other channels contribute very little ($-0.058$).
This is consistent with early experiments which demonstrated that 
the contribution of non-universal three-body contact to the momentum distribution appears to be {\em unmeasurable} near resonance ~\cite{Wild12,comment}. In Ref.~\cite{Wild12}, the authors measured Tan's contact using rf spectroscopy for $^{85} \mathrm{Rb}$ atoms. It is demonstrated that when fitted to the frequency dependence
of the tail of the rf spectrum, the experiment data exhibit no visible
evidence of measurable three-body effects.
Equally importantly, Eq.~\eqref{universal} shows that in general the three-body and other higher order effects (i.e. the terms proportional to $\epsilon$) become
more important when the dimensionality decreases. This is again fully consistent with the previous renormalization studies which show
that the three-body effect increases from a few percent in 3D~\cite{Borzov12} to around $20\% \sim 40\%$ in 2D~\cite{Mashayekhi13}.

In conclusion, we have obtained a rigorous solution to a unitary Bose gas
or a quantum gas at infinite scattering length, which offers an answer to
one of the long-standing challenges in quantum many-body physics. This
solution can further shed light on future studies of other aspects of
large-scattering-length physics such as the ultra-violet properties of a Bose gas.

This work is in part supported by CIFAR, NSERC (Canada), NKBRSFC under
grants No.~2011CB921502 and No.~2012CB821305, and NSFC under grants
No.~11228409, No.~61227902, and No.~61378017. We thank Joseph
H. Thywissen, 
Tao Xiang, Yupeng Wang, and Zhengyu Weng  for constructive discussions.

{\it Note added.---}After the submission of our manuscript, we noticed a new experimental work on universal Bose gases ~\cite{Makotyn14}, and a subsequent analysis of the data obtained in the above mentioned experiment~\cite{Smith}.


\begin{thebibliography}{References}
\bibitem{dilute-limit}For $N = 4,6,...$-body forces, they in general read as
$	A^{(N)} =i c_1^{(N)} \epsilon + a_1^{(N)} \epsilon (k_{\mu} \lambda_B)^{-2(N-3) -
		\epsilon}+...$
Contributions from $a_2^{(N)}$  to the energy density simply contain higher
powers of $\epsilon$ and can be ignored in the $\epsilon$ expansion.
Meanwhile, the contributions of $A^{(3)}$ are of the form 
$	A^{(3)} =i c_1^{(3)} \epsilon+ a_2^{(3)} \epsilon^2	+ ...$,
where $c_1^{(3)}$ is the imaginary part of the one-loop 3-body contribution and
gives the most dominating contribution to
the imaginary part of the ground state energy density; the next order terms are
from $L=2$ processes and contain higher powers of $\epsilon$.
Similarly, $A^{(5,7,...)}$ have the form,
$	A^{(N)} =i c_1^{(N)} \epsilon + a_2^{(N)} \epsilon^2 (k_{\mu}
	\lambda_B)^{-2(N-4)-2 \epsilon} +...$
\bibitem{Energy_at_Res} The ground state energy density up to the 1-loop effect can be summed as 
$E(n_0, \mu) = \frac{g_2 n_0^2}{2} \frac{1}{1-(-i k_{\mu} \lambda_B)^{2-\epsilon}} - \frac{i}{2} \int \frac{d \omega}{2 \pi} \frac{d^d k }{(2 \pi)^d} \ln \left( 1+ \frac{g_2^2 n_0^2}{[1-(-i k_{\mu} \lambda_B)^{2-\epsilon}]^2} G( \omega,k) G(-\omega, -k) \right) - \frac{g_2^2 n_0^2}{2[1-(-i k_{\mu} \lambda_B)^{2-\epsilon}]^2} \int \frac{d^d k}{(2 \pi)^d} \frac{1}{-k^2 + k_{\mu}^2 + i \delta} $, where $G(\omega, k) = i(\omega - k^2/2 + k_{\mu}^2/2 - \Sigma(\omega,k) + i \delta)^{-1}$ is the dressed Green's function and $\Sigma(\omega,k) = \frac{2 g_2 n_0} {1- \left( \lambda_B \sqrt{k^2/4 - \omega - k_{\mu}^2 - i \delta}\right)^{2-\epsilon}}$ is the self energy.
Taking the derivatives with respect to $n_0$ and $\mu$ and substituting into Eq.~(\ref{SCD}), we obtain the self-consistent equation below.
\end{thebibliography}
\end{document}